\documentclass{article}
\usepackage{LaThuileFPSpro_kg}
\input{colordvi.sty}\newcommand {\pom} {I\!\! P}
\begin{document}
\title{~\\~\\~\\~\\~\\~\\~\\~\\~\\~\\~\\~\\~\\~\\
HADRONIC DIFFRACTION: WHERE DO WE STAND?\footnote{\hspace{0.2cm}To appear in: 
{\em Proc. Les Rencontres de Physique de la Vall\'{e} d'Aoste}, La Thuile, Aosta 
Valley, Italy, February 29 - March 6, 2004.}}
\author{
  Konstantin Goulianos        \\
  {\em The Rockefeller University, 1230 York Avenue, New York, NY 10021, USA}}
\maketitle

\baselineskip=11.6pt
\begin{abstract}
Experimental results on hadronic soft and hard diffractive processes 
are reviewed 
with emphasis on aspects of the data that point to the underlying QCD 
mechanism for diffraction. Diffractive differential cross sections 
are shown to be factorized into two terms, 
one representing the total cross section at the reduced energy, 
corresponding to the rapidity region(s) in which there is particle 
production, and another interpreted as the probability 
of formation of the rapidity gap(s) characterizing diffraction. 
By (re)normalizing the term of gap formation probability to unity,
cross sections for single, central, and multiple rapidity gap 
soft diffraction, as well as structure functions 
for hard diffraction processes,
are obtained from the underlying inclusive parton distribution functions.
A unified partonic picture emerges, in which 
diffraction appears to be mediated by the exchange of low-$x$ partons 
subject to color constraints. 
\end{abstract}
\begin{center}
\begin{figure}[h]
\end{figure}
\end{center}
\newpage
\section{Introduction}
Hadronic diffraction has traditionally been treated in the framework of 
Regge theory\cite{Collins}. 
In this approach, the key 
player mediating diffractive processes is the Pomeron 
($\pom$) {\em Regge trajectory}, 
presumed to be delineated by a ``family'' of particles carrying 
the quantum numbers of the vacuum. Although no particles 
were known (and have yet to be found!) to belong to this family, 
the Pomeron trajectory 
was introduced in the 1970s to account for the observations that 
the $K^+p$ cross section was increasing with energy at the Serpukov 
70 GeV proton synchrontron, and the elastic and total $pp$ cross sections, 
which were falling with increasing energy, started to flatten out and then 
increase as larger collision energies became available
at the CERN Intersecting Storage Ring (ISR) $pp$ collider.

Regge theory worked reasonably well in describing elastic, 
diffractive and total hadronic cross sections at energies up to 
$\sqrt s\sim 50$ GeV, spanning the range of Serpukov, Fermilab, and
ISR energies.  All processes could be accommodated 
in a simple Pomeron pole approach. This success was documented, 
among other places, in a 1983 Physics 
Reports article by this author\cite{physrep}.    
Results from a Rockefeller University experiment on photon 
dissociation on hydrogen published in 1985\cite{Chapin} were also 
well described in this approach. 

The early success of Regge theory was precarious. 
The theory was known to asymptotically violate unitarity, 
as the $\sim s^{\epsilon}$ power law increase of total cross sections 
would eventually exceed the
Froissart bound of $\sigma_T<C\cdot \ln^2s$ based 
on analyticity and unitarity.
But the confrontation with unitarity came far 
earlier than what would be considered asymptopia by Froissart bound 
considerations. 
As collision energies climbed upwards in the 1980s to reach $\sqrt s=630$ 
GeV at the CERN S$\bar pp$S collider and $\sqrt s=1800$ GeV at the 
Fermilab Tevatron $\bar pp$ collider, diffraction dissociation 
could no longer be accommodated within a factorizable 
Regge pole framework in which the Pomeron exchange contribution to 
total, elastic, and single diffractive $pp$  
cross sections is given by
\begin{equation}
\sigma^{tot}(s)=\beta^2_{\pom pp}(0)
\left(\frac{s}{s_0}\right)^{\alpha_{\pom}(0)-1}
\label{eq:total}
\end{equation}
\begin{equation}
\frac{d\sigma^{el}}{dt}=\frac{\beta^4_{\pom pp}(t)}{16\pi}\;
{\left(\frac{s}{s_0}\right)}^{2[\alpha_{\pom}(t)-1]}
\label{eq:elastic}
\end{equation}
\begin{equation}
\frac{d^2\sigma_{sd}}{d\xi dt}=
\underbrace{\frac{{\beta_{\pom pp}^2(t)}}{16\pi}\;\xi^{1-2\alpha_{\pom}(t)}}_
{f_{\pom/p}(\xi,t)}
\left[\beta_{\pom pp}(0)\,g(t)
\;\left(\frac{s'}{s_0}\right)^{\alpha_{\pom}(0)-1}\right]
\label{eq:diffractive}
\end{equation}

\noindent where $\alpha_{\pom}(t)=
\alpha_{\pom}(0)+\alpha' t=(1+\epsilon)+\alpha' t$
is the Pomeron
trajectory, $\beta_{\pom pp}(t)$ the coupling of the Pomeron to the proton,
$g(t)$ the $\pom\pom\pom$  coupling, $s'=M^{2}$ the $\pom-p$ center of
mass energy squared, $\xi = 1-x_{F}=s'/s=M^2/s$ the fraction of
the momentum of the proton carried by the Pomeron, and $s_0$ an energy
scale parameter traditionally set to the hadron mass scale of 1~GeV$^2$.

The single diffractive cross section, Eq.~(\ref{eq:diffractive}), factorizes 
into two terms, the one in square brackets, which can be viewed as the 
$\pom$-$p$ total cross section, and the other labeled $f_{\pom/p}(\xi,t)$, 
which may be interpreted as the Pomeron flux emitted by the diffracted proton. 
In 1985, Ingelman and Schlein (IS) proposed that the Pomeron may have 
partonic structure and, assuming various forms for its structure, 
predicted diffractive dijet production rates by replacing the $\pom$-$p$ 
total cross section in Eq.~(\ref{eq:diffractive}) by the  
parton level cross section. When later the  UA8 Collaboration discovered 
and characterized diffractive dijet production in $\bar pp$ collisions at 
$\sqrt s =630$~GeV\cite{UA8}, the reported measured rate was severely 
suppressed relative to that expected from the IS model using 
a Pomeron flux based on Regge theory and factorization.  
However, it was not clear at that time whether this 
discrepancy represented a breakdown of factorization or 
failure of the IS model. 

The first clear experimental evidence for 
a breakdown of factorization in Regge theory was 
reported by the CDF Collaboration in 1994\cite{CDF_sd}. CDF measured 
the single diffractive cross section in $\bar pp$ collisions 
at $\sqrt s=$546 and 1800 GeV and found a suppression factor 
of about an order of magnitude at $\sqrt s=$1800 GeV relative to 
predictions based on extrapolations from $\sqrt s\sim$20 GeV. 
The suppression factor at $\sqrt s=$546 GeV was $\sim 5$, 
approximately equal to that reported by UA8 for hard diffraction 
at $\sqrt s=$630 GeV. The near equality of the suppression in soft and 
hard diffraction provided the clue that led to the development 
by this author of a {\em renormalization} procedure for single 
diffraction, which was later extended to central and multigap 
diffractive processes, as described in the following sections. 

\section{Renormalization and scaling in single diffraction\label{sec:renorm}}
The breakdown of factorization in Regge theory was traced to 
the energy dependence of $\sigma_{sd}^{tot}(s)\sim s^{2\epsilon}$, 
which is faster than that of $\sigma^{tot}(s)\sim s^\epsilon$,
so that as $s$ increases unitarity would have to be 
violated if factorization holds. 
This is reflected in an explicit $s$-dependence in $d\sigma_{sd}(M^2)/dM^2$:
\begin{equation}
\hbox{ Regge theory: }
d\sigma_{sd}(M^2)/dM^2\sim s^{2\epsilon}/(M^2)^{1+\epsilon}
\label{reggeM2}
\end{equation}

\begin{center}
\begin{figure}[h]
\includegraphics{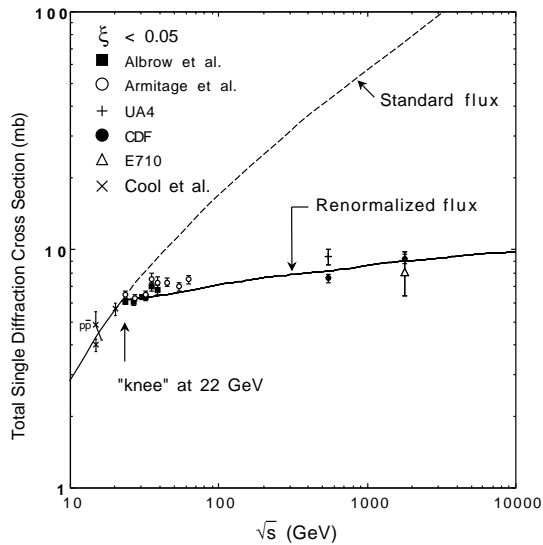}
\vglue 7.5cm
\caption{\it  Total \protect{$pp/\bar pp$}
single diffraction dissociation cross section data (\,{\em both} $\bar p$ and $p$ sides)   
for \protect{$\xi<0.05$} compared with predictions 
based on the standard and the renormalized Pomeron flux [from Ref.\protect\cite{R}].}
\label{fig:R}
\end{figure}
\end{center}

In a paper first presented by this author in 1995  
at La Thuile\cite{lathuile95} and Blois~\cite{blois95}
and later published in Physics Letters~\cite{R},
it was shown that unitarization could be achieved, and the factorization 
breakdown in single diffraction fully accounted for, by 
interpreting the Pomeron flux 
of Eq.~(\ref{eq:diffractive}) as a probability density and {\em renormalizing} 
its integral over $\xi$ and $t$ to unity, 
\begin{eqnarray}
\hbox{ renormalization: }f_{\pom/p}(\xi,t)\Rightarrow N_s^{-1}\cdot f_{\pom/p}(\xi,t)\\ 
\nonumber\\
\hbox{where } N_s\equiv \int_{\xi(min)}^{\xi(max)}d\xi\int_{t=0}^{-\infty}dt\,f_{\pom/p}(\xi,t)\sim s^{2\epsilon}\nonumber
\label{eq:renorm}
\end{eqnarray}
\noindent where $\xi(min)=M_0^2/s$ (with $M_0^2=1.4$ GeV$^2$: effective threshold for diffraction dissociation), and $\xi(max)=0.1$. 
The energy dependence of $N^{-1}_s$, introduced by renormalization, removes 
the explicit $s$-dependence from $\sigma_{sd}^{tot}$, thereby ensuring 
unitarization. In Fig.~\ref{fig:R}, $\sigma^{tot}_{sd}(s)$ is compared with 
Regge predictions using the standard or renormalized Pomeron flux. The
renormalized prediction is in excellent agreement with the data.   

\begin{center}
\begin{figure}[h]
\includegraphics{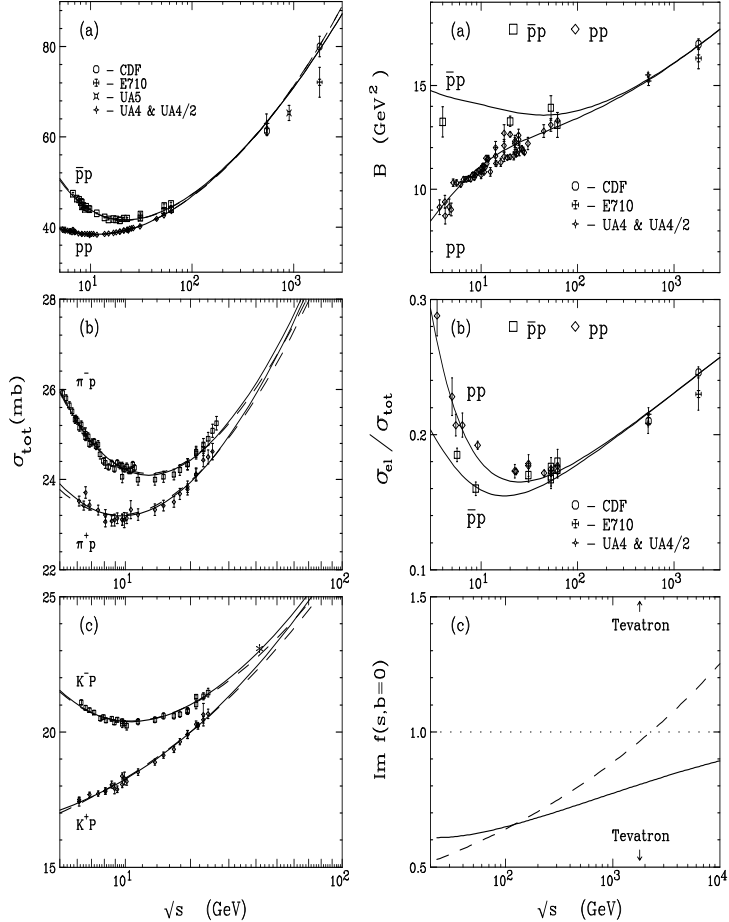}
\vglue 12.3cm
\caption{\it  Fits to total cross section and elastic scattering data. The dashed lines are Born level Regge fits, and the solid lines are fits after eikonalization of the elastic scattering amplitude [from Ref.\protect\cite{CMG}].} 
\label{fig:CMG}
\vglue -1cm
\end{figure}
\end{center}

It should be noted that the elastic and total cross sections are not 
affected by this procedure. One way to achieve unitarization 
in these cases is by using the eikonal approach, as reported by
Covolan, Montanha and this author\cite{CMG}. As shown in Fig.~\ref{fig:CMG},
excellent agreement is obtained between elastic and total 
cross section data and predictions based on Regge theory and 
eikonalization. 

\begin{center}
\begin{figure}[h]
\includegraphics{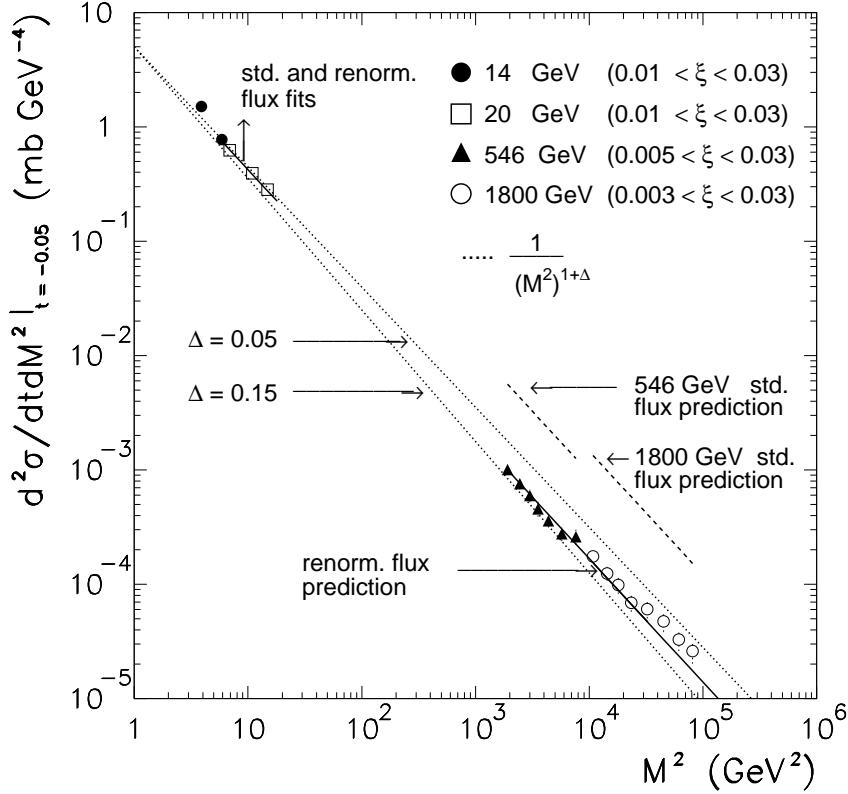}
\vglue 10cm
\caption{\it  Cross sections \protect$d^2\sigma_{sd}/dM^2 dt$
for $p+p(\bar p) \rightarrow p(\bar p)+X$ at
$t=-0.05$ GeV$^2$ and $\protect\sqrt s=14$, 20, 546 and 1800 GeV.
Standard (renormalized) flux predictions
are shown as dashed (solid) lines.
At $\protect\sqrt s$=14 and 20 GeV,
the fits using the standard and renormalized fluxes coincide
 [from Ref.\protect\cite{GM}].}
\label{fig:GM}
\end{figure}
\end{center}

An important aspect of renormalization is that it leads to a scaling 
behavior, whereby $d\sigma_{sd}(M^2)/dM^2$ has no explicit $s$-dependence: 
\begin{equation}
\hbox{ $M^2$-scaling: }d\sigma_{sd}(M^2)/dM^2\sim 1/(M^2)^{1+\epsilon}
\label{eq:renormM2}
\end{equation} 
\noindent  This `scaling law' was demonstrated for {\em differential} 
soft single diffractive cross sections in a paper by Montanha 
and this author\cite{GM},
in which renormalization was exploited to achieve a 
fit to all $pp$ and $\bar pp$ cross section data 
with only one free parameter -  
see also Refs.\cite{frascati,d2000sf,d2000hadron,perspective,cipanp} 
for scaling behavior in hard diffraction. 
Figure \ref{fig:GM} shows a comparison 
between data and the standard Regge and renormalization predictions for  
$d\sigma_{sd}(M^2)/dM^2$. The renormalization prediction provides an 
excellent fit to the data over five orders of magnitude. 

As discussed below, $M^2$-scaling 
represents a general behavior extending to central and 
multigap diffractive processes.

\section{Central rapidity gaps: double diffraction}
Double diffraction dissociation is the process in which both 
colliding hadrons dissociate leading to events with a central 
rapidity gap{\footnote{We use rapidity, 
$y\equiv\frac{1}{2}\ln\frac{E+p_L}{E-p_L}$, and pseudorapidity, 
$\eta\equiv -ln(\tan\frac{\theta}{2})$, interchangeably, as they are 
approximately equal in the kinematic range of interest.
} (see Fig.~\ref{fig:dd}). 

\begin{center}
\begin{figure}[h]
\includegraphics{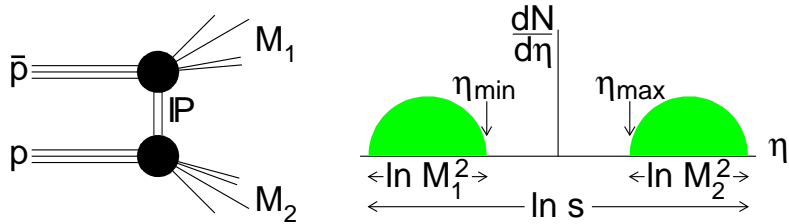}
\vglue 3cm
\caption{\it  Schematic diagram and event topology 
of $\bar pp$ double diffraction dissociation; the shaded areas 
represent regions of particle production [from Ref.\protect\cite{CDF_dd}].} 
\label{fig:dd}
\end{figure}
\end{center}
\vglue -0.3cm
In Regge theory, the DD
cross section is given by\cite{physrep}
\begin{equation}
{d^{3}\sigma^{dd}\over dtdM_1^2dM_2^2}  =  
  {d^{2}\sigma_1^{sd}\over dtdM_1^2}{d^{2}\sigma_2^{sd}\over dtdM_2^2}
  \,/\,{d\sigma^{el}\over dt} 
   ={[\kappa \beta_1(0)\beta_2(0)]^2\over 16\pi}\,
  {{s^{2\epsilon}\, e^{b_{dd}t}\over (M_1^2M_2^2)^{1+2\epsilon}}}
\label{eq:ddM2}
\end{equation}
where $b_{dd}=2\alpha'\ln{({\textstyle s}{\textstyle s_{\circ}}/M_1^2M_2^2)}$.
Since there is no term in this formula that can be identified as Pomeron 
flux, it was not immediately clear how renormalization should be applied 
in DD. The procedure employed in Ref.\cite{R} is, in hindsight, 
incorrect. The clue to the correct procedure is embedded in 
a 1998 paper by this author\cite{brazil},
in which the SD cross section is expressed in terms of 
the rapidity gap variable, $\Delta\eta$, in place of $\xi$ 
or $M^2$. These variables are related by
\begin{equation}
\Delta\eta=\ln \frac{s}{M^2}=-\ln\xi
\label{rapgap}
\end{equation}
Using rapidity gap variables, the SD and DD cross sections take the forms
\begin{eqnarray}  
{d^{2}\sigma_{sd}\over dtd\Delta \eta}&=& 
  \left[~{\beta^{2}(t)\over 16\pi} e^{2[\alpha(t)-1]\Delta \eta}~\right]
  \cdot\kappa\left[\beta^{2}(0)
{\left(\frac{s'}{\textstyle s_{\circ}}\right)}^{\epsilon}\right] 
\\
{d^{3}\sigma_{dd}\over dtd\Delta \eta d\eta_c}&=&
\left[{\kappa\beta^{2}(0)\over 16\pi} e^{2[\alpha(t)-1]\Delta \eta}\right]
\cdot\kappa\left[\beta^{2}(0){{\left(\frac{s'}{\textstyle s_{\circ}}\right)}}^
{\epsilon}
\right] 
\label{eq:sddd}
\end{eqnarray}
where $\kappa\equiv g_{\pom\pom\pom}/\beta_{\pom pp}$,
$\eta_c$ the center of the rapidity gap, and $s'$ the reduced collision 
energy squared, defined by 
\begin{equation}
\hbox{ reduced energy: }\ln \frac{s'}{s_0}=\sum_i\ln \frac{M^2_i}{s_0}~\Rightarrow~\frac{s'}{s_0}=\exp\left[{\sum_i\Delta\eta'_i}\right]
\label{eq:reduced} 
\end{equation}
The expressions for the SD and DD cross sections 
are strikingly similar, except that in DD
the gap is not ``tied'' to the (anti)proton and therefore 
$\eta_c$ is treated 
as an independent variable. The two factors in brackets on the 
right hand side are identified as the rapidity gap probability 
and the reduced energy total cross section, respectively.
In both SD and DD, the gap probability has the same $\Delta\eta$ dependence 
and its integral over all phase space is $\sim s^{2\epsilon}$: 
\begin{equation}
P_{gap}(\Delta\eta)\sim e^{2\epsilon\Delta\eta}~~~\Rightarrow~~~
\int_{\Delta\eta(min)}^{\ln s}P_{gap}(\Delta\eta)\,d\Delta\eta\sim s^{2\epsilon} 
\label{eq:gap}
\end{equation}
Thus, renormalization cancels the $s^{2\epsilon}$ factor 
in Eq.~(\ref{eq:ddM2}), ensuring $M^2$-scaling and predicting 
similar suppression factors for SD and DD. In 2001, CDF reported results 
verifying this prediction\cite{CDF_dd}. Fig.~\ref{fig:ddtot}
shows a comparison between DD total cross sections versus $s$ 
and predictions based on Regge theory with and without renormalization.
The renormalized prediction is in excellent agreement with the data.
\begin{center}
\begin{figure}[h]
\includegraphics{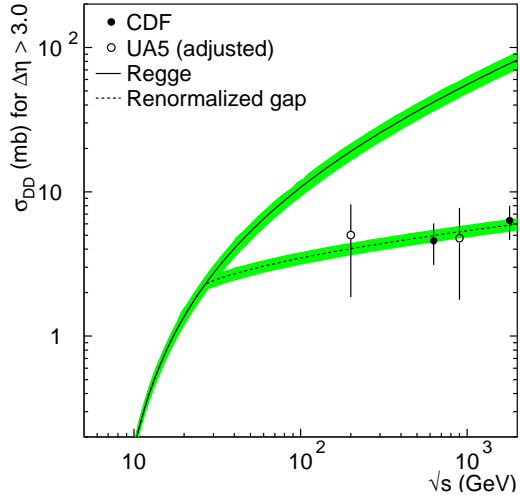}
\vglue 6.5cm
\caption{\it  The total $\bar pp$ double diffractive cross section 
versus $s$ [from Ref.\cite{CDF_dd}].}
\label{fig:ddtot}
\end{figure}
\end{center}
\newpage
\section{Multigap diffraction\label{sec:multigap}}
Following the success of the renormalized gap probability model in 
correctly predicting double diffraction dissociation,
the next challenge was to understand events with multiple rapidity gaps. 
The classical example of a two-gap process is 
double Pomeron exchange, $\bar p+p\rightarrow \bar p+(gap)+X+(gap)+p$, where 
the $\bar p$ and $p$ are scattered quasi-elastically and a central system 
$X$ of mass $M$ is produced separated from the outgoing nucleons by large 
rapidity gaps (see Fig.~\ref{fig:idpe}). 
\begin{center}
\begin{figure}[htp]
\includegraphics{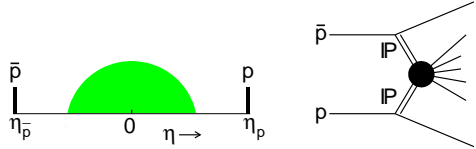}
\vglue 1.5cm
\caption{\it  DPE: Double Pomeron Exchange} 
\label{fig:idpe}
\end{figure}
\end{center}
In Regge theory, the expression for DPE analogous to that of 
Eq.~(\ref{eq:diffractive}) for SD is
\begin{equation}
\frac{d^4\sigma}{dt_{\bar p}dt_pd\xi_{\bar p}d\xi_p}=f_{\pom/\bar p}(\xi_{\bar p},t_{\bar p})\cdot f_{\pom/p}(\xi_p,t_p)\cdot
\kappa^2\left[\beta^2(0)\left(\frac{s'}{s_0}\right)^\epsilon\right]
\label{eq:idpe}
\end{equation}
When the renormalization model was proposed\cite{R}, it was presumed 
that both the $\bar p$ and $p$\, Pomeron fluxes ought to be 
renormalized to unity. This leads to a DPE cross section
which is doubly suppressed as $\sigma_{sd}$.  
After generalizing the Pomeron flux model it became 
clear that the gap probability 
depends on the sum of the two gaps and therefore there should be no additional 
suppression due to the second gap. In 2001, a 
procedure for multigap renormalization was proposed by 
this author\cite{BNL,nutsandbolts,corfu} on the basis of which 
explicit predictions were made for DPE as well as for a similar 
two-gap process  in which the proton dissociates. 
The latter, named SDD (for SD + DD),
leads to SD events with a central rapidity gap within the diffraction 
dissociation cluster (see Fig. \ref{fig:sdd}).
\begin{center}
\begin{figure}[htp]
\includegraphics{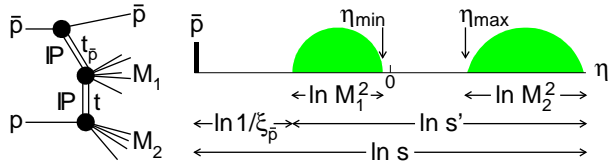}
\vglue 1.9cm
\caption{\it  SDD: Single+Double Diffraction [from Ref.\cite{CDF_sdd}].} 
\label{fig:sdd}
\end{figure}
\end{center}
In 2003, CDF tested the renormalization model 
predictions for the above two-gap processes\cite{CDF_idpe,CDF_sdd}, 
reporting agreement between data and model both in shape and normalization. 
In Fig. \ref{fig:gapratios}, data are 
compared with theory for one-gap to no-gap and 
two-gap to one-gap cross section ratios. While the one-gap/no-gap 
ratio is severely suppressed relative to the Regge theory 
prediction, the two-gap/one-gap ratios are relatively non-suppressed 
and equal to $\approx \kappa$. This result is expected in the renormalization 
model, since two-gap and one-gap cross sections are 
$\propto \kappa^2$ and $\propto \kappa$, 
respectively, and gap probability factors are normalized to unity.  
In the next section we examine how the scaling features of 
multigap diffraction arise naturally in a QCD based framework for diffraction.

\begin{minipage}[t]{5.5cm}
\includegraphics{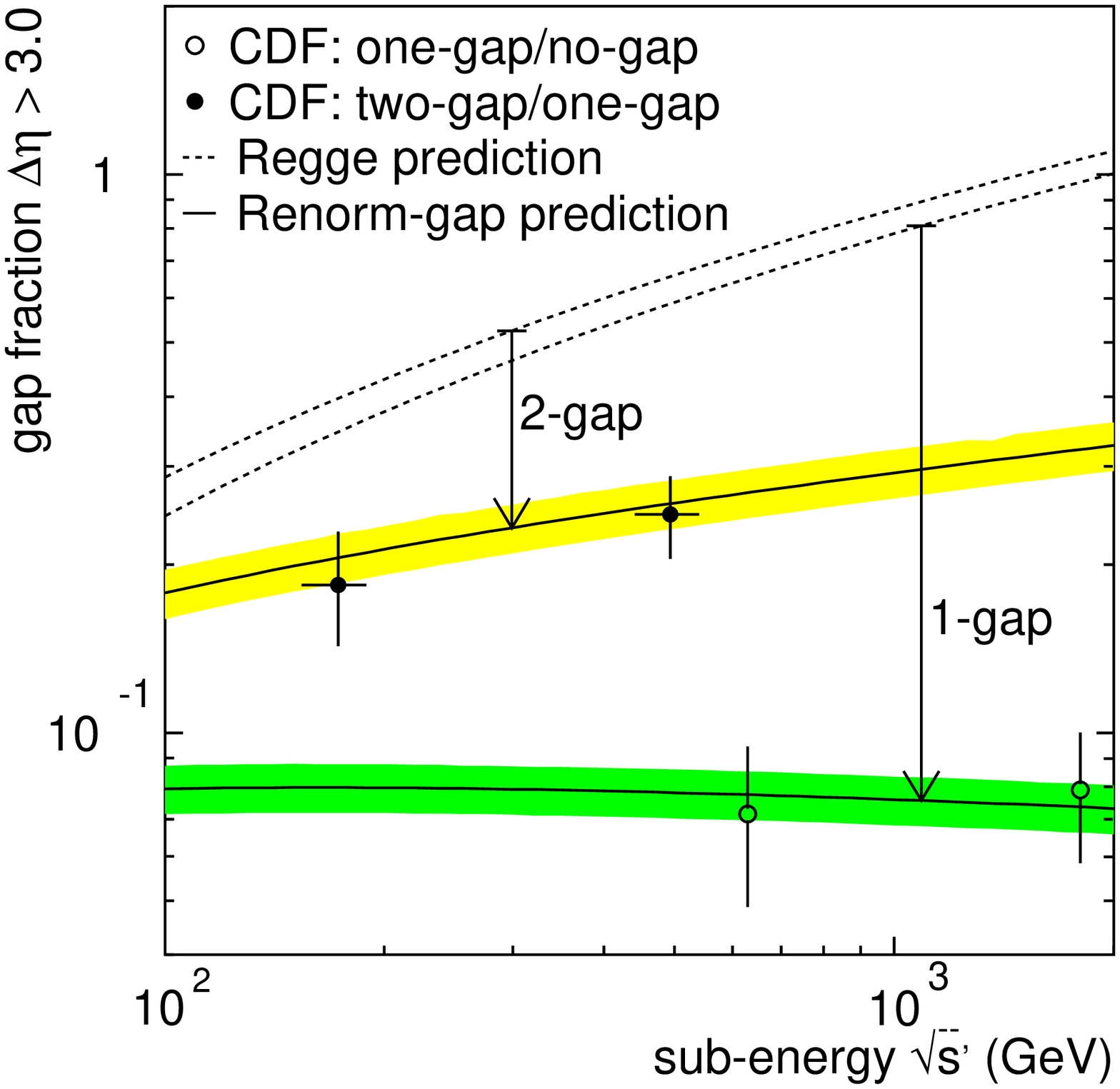}
\hfill
\end{minipage}
\begin{minipage}[t]{5.5cm}
\includegraphics{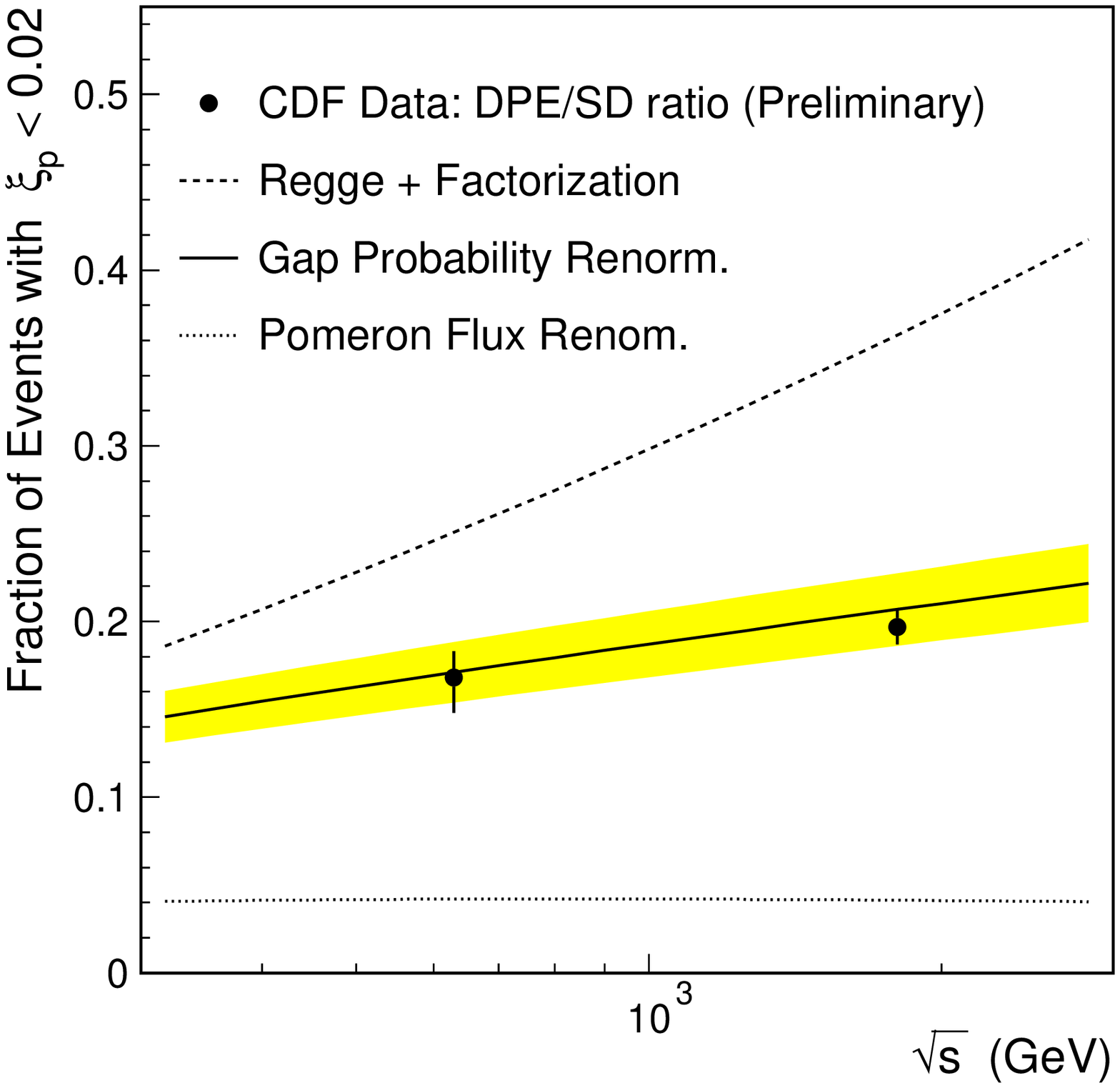}
\end{minipage}
\vglue 5.5cm
\begin{figure}[h]
\caption{\it  Comparison of ratios of cross sections with  
standard Regge theory and renormalization predictions: 
({left}) ratios of SDD to SD (two-gap/one-gap) and SD to ND (one-gap/no-gap);
({right}) ratio of DPE to SD (two-gap/one-gap).}
\label{fig:gapratios}
\end{figure}  

\section{A parton model approach to diffraction~\label{approach}}
The form of the rise of total cross sections at high energies, 
$\sim s^\epsilon$, which in Regge theory requires a Pomeron trajectory 
with intercept $\alpha(0)=1+\epsilon$, is the form expected  
in a parton model approach, where cross sections are 
proportional to the number of available wee partons\cite{Levin}.
In terms of the rapidity region in which there is particle 
production\footnote{We assume $p_T=1$ GeV so that $\Delta y'=\Delta \eta'$.}, 
$\Delta \eta'$, the total $pp$ cross section is given by
\begin{equation}
\sigma_{pp}^{tot}=\sigma_0\cdot e^{\epsilon\Delta\eta'}
\label{totDeta}
\end{equation}
Since from the optical theorem the total cross section is proportional 
to the imaginary part of the forward ($t=0$) 
elastic scattering amplitude, the full parton 
model amplitude may be written as 
\begin{equation}
{\rm Im\,f^{el}}(t,\Delta\eta)\sim e^{({\epsilon}+\alpha't)\Delta \eta}
\label{eq:fPM}
\end{equation}
\noindent where we have added to $\epsilon$ the term  $\alpha'(t)$ as 
a parameterization of the $t$-dependence of the amplitude. 

Based on this amplitude, the diffractive cross sections corresponding to the 
gap configurations we discussed above are expected to have the forms
\begin{eqnarray}
\frac{d^2\sigma_{sd}}{dt\,d\Delta\eta}=N^{-1}_{gap}(s)& 
F_p(t)\left\{e^{[\epsilon+\alpha'(t)]\Delta\eta}\right\}^2
 &\kappa\left[\sigma_0e^{\epsilon\Delta\eta'}\right]
\nonumber\\
\frac{d^3\sigma_{dd}}{dt\,d\Delta \eta\,d\eta_c}=N^{-1}_{gap}(s)&
\left\{e^{[\epsilon+\alpha'(t)]\Delta\eta}\right\}^2
&\kappa\left[\sigma_0e^{\epsilon(\Sigma_i\Delta\eta_i')}\right]
\nonumber\\
\frac{d^4\sigma_{sdd}}{dt_1\,dt_2\,d\Delta \eta\,d\eta_c}=N^{-1}_{gap}(s)&
F_p(t)\Pi_i\left\{e^{[\epsilon+\alpha'(t_i)]\Delta\eta_i}\right\}^2
&\kappa^2\left[\sigma_0e^{\epsilon(\Sigma_i\Delta\eta_i')}\right]
\nonumber\\ 
\frac{d^4\sigma_{dpe}}{dt_1\,dt_2\,d\Delta \eta\,d\eta'_c}=N^{-1}_{gap}(s)&
\underbrace{\Pi_{i}\left\{F_p(t_i)e^{[\epsilon+\alpha'(t_i)]\Delta\eta_i}\right\}^2}_{\hbox{gap probability factor}} 
&\kappa^2\underbrace{\left[\sigma_0e^{\epsilon(\Delta\eta')}\right]}_{\textstyle{\sigma^{tot}(s')}}
\label{eq:diffPM}
\end{eqnarray}
where, as in Eq.~(\ref{eq:gap}), the (re)normalization factor 
$N_{gap}(s)$ is the integral of the gap probability factor over all 
phase space in $t_i$, $\Delta\eta_i$ and the variables $\eta_c$ and 
$\eta'_c$, which represent the center of the ``floating'' (not adjacent to 
a nucleon) rapidity gap in DD or SDD and the floating cluster in DPE, 
respectively. In each case, the independent variables are the ones 
on the left hand side of the equation, but for pedagogical reason  we 
use on the right hand side additional variables, which can  
be expressed in terms of the ones on the left. 

A remarkable property of the above expressions is that they factorize 
into two term, one depending on the rapidity region(s) in which there is 
particle production and the other that consists of the rapidity gap(s). 
This is due to the exponential dependence 
on $\Delta\eta$ of the elastic amplitude, which allows non-contiguous 
regions in rapidity to be added in the exponent. A consequence of this
is that the (re)normalization factor is $\sim s^{2\epsilon}$ in all cases, 
ensuring universal $M^2$-scaling.

These expressions may be understood as follows:
(i) the term in square brackets represents the nucleon-nucleon 
total cross section at the reduced energy defined in 
Eq.~(\ref{eq:reduced}); (ii) the factors $\kappa$, one for each gap, 
are the color factors required to enable rapidity gap formation; (iii)
the gap probability factor is the amplitude squared of the elastic scattering 
between a diffractively dissociated and a surviving proton, in which case 
it contains the proton form factor, $F_p(t)$, or between two diffractively 
dissociated protons.
Since the reduced energy cross section is properly normalized, the 
gap probability term is (re) normalized to unity using the $N^{-1}_{gap}$ factor. 

\subsection{The parameters $\textstyle\epsilon$ and $\textstyle\kappa$}
The parameters $\epsilon$ and $\kappa$ in Eqs.~(\ref{eq:diffPM}) have been 
experimentally found to be\cite{CMG,GM}
\begin{eqnarray}
\hbox{ experiment:       }\epsilon\equiv &\alpha_{\pom}(0)-1&=0.104\pm 0.002\pm 0.01\hbox{ (syst)}\nonumber\\
\kappa\equiv &\frac{\textstyle{g}_{\pom\pom\pom}}{\beta_{\pom p}}&=0.17\pm0.02\hbox{ (syst)}
\label{eq:ek_exp}
\end{eqnarray} 
where the systematic error assigned to $\epsilon$ is a rough 
estimate by this author 
based on considering fits made to cross section data by various authors.

\begin{minipage}[t]{5.5cm}
\includegraphics{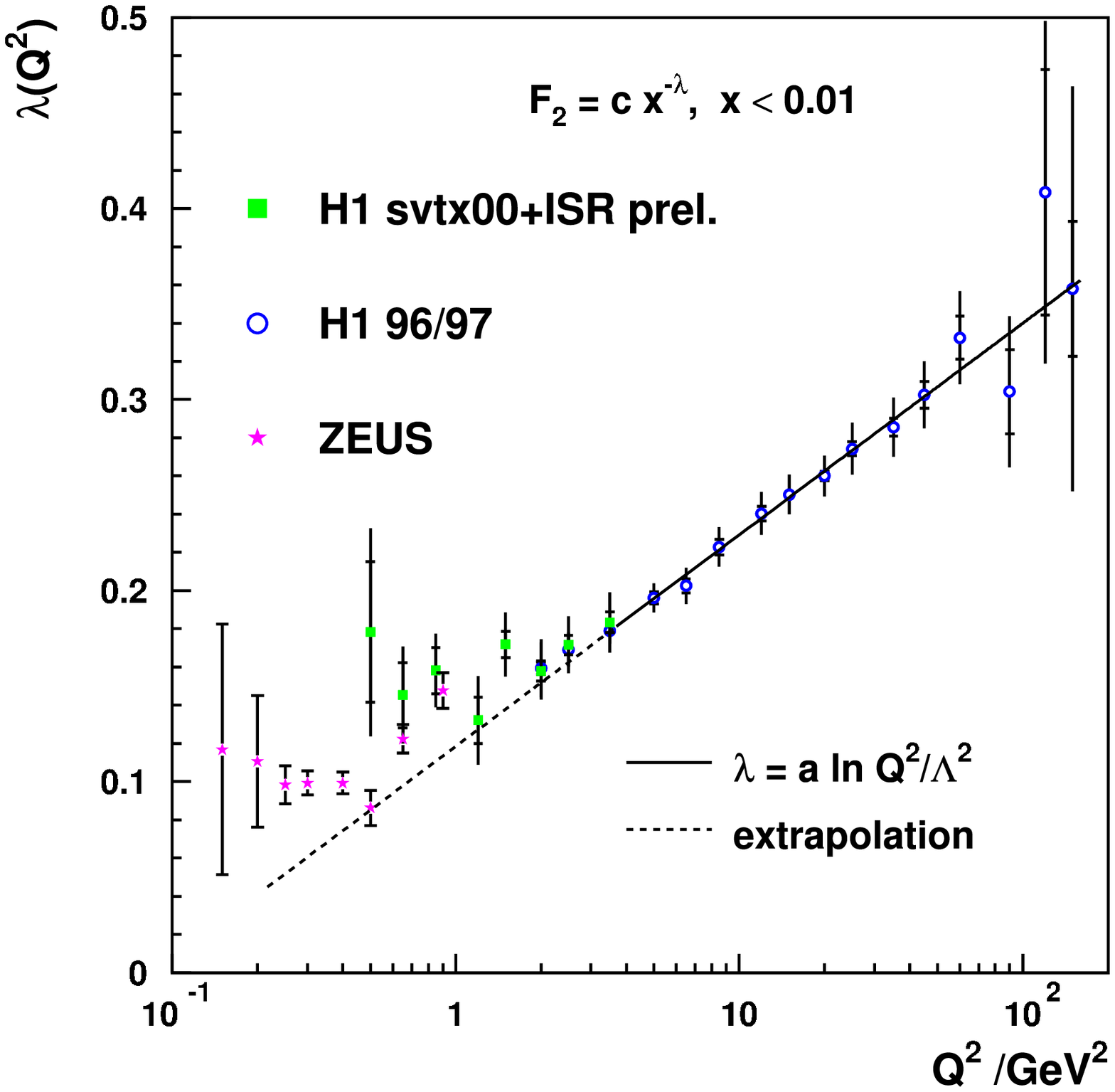}
\hfill
\end{minipage}
\begin{minipage}[t]{5.5cm}
\includegraphics{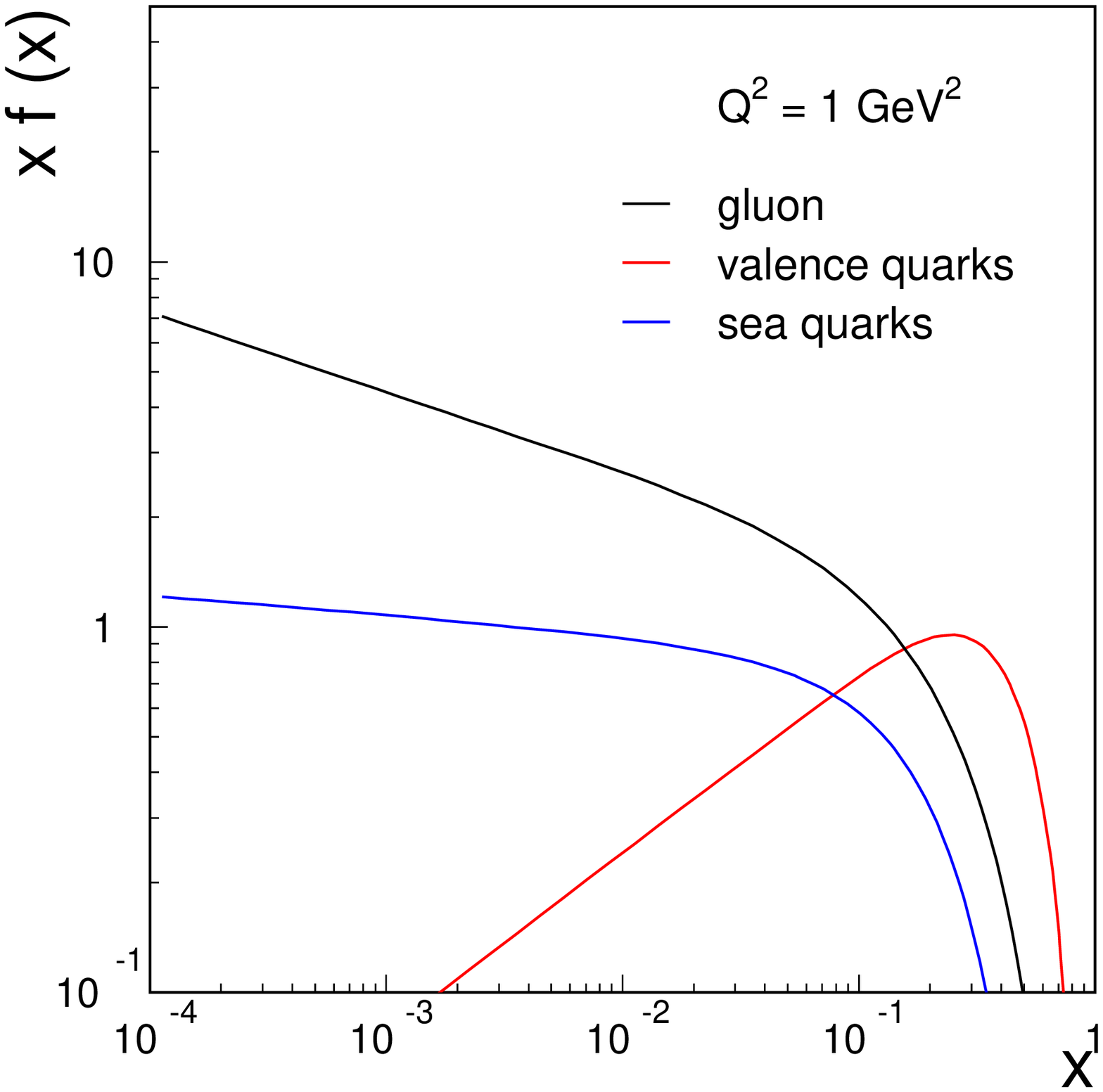}
\vglue 1.2cm
\hspace {1.8cm}\hspace{1cm}$\lambda_g=0.2$\\
\vglue 0.5cm
\hspace {1.8cm}\hspace{1cm}$\lambda_q=0.04$\\
\vglue 1.1cm
\hspace {1.8cm}\hspace{1cm}$\lambda_R=-0.5$\\
\end{minipage} 
\begin{center}
\begin{figure}[htp]
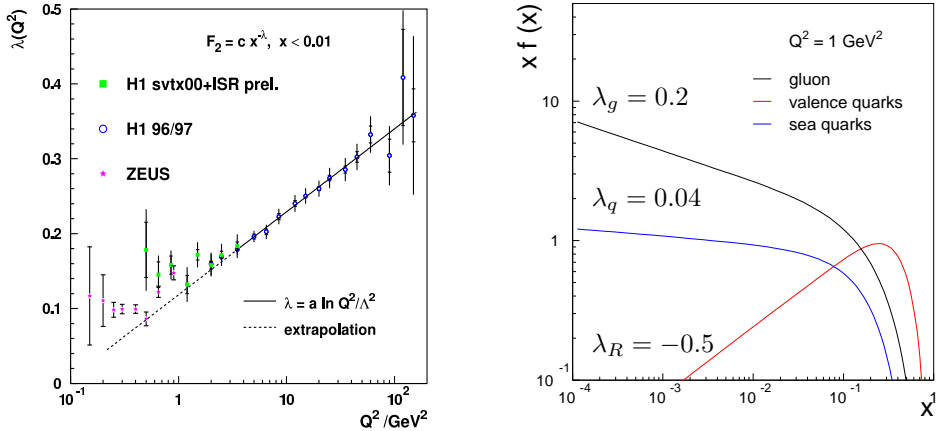

\caption{\it  ({left}) The parameter $\lambda$ versus $Q^2$ 
of a fit to the structure function $F_2(x,Q^2)\sim x^{-\lambda}$ in DIS at HERA\cite{H1_lambda};
({right}) CTEQ5L nucleon parton distribution functions for $Q^2=1$ GeV$^2$.} 
\label{fig:softQCD}
\end{figure}
\end{center}
Measurements of parton densities at HERA indicate that partonic 
structure in the nucleon may exist down to the 
hadron mass scale of $Q^2\approx 1$ GeV$^2$. 
This is seen in Fig.~\ref{fig:softQCD} (\em left}), where the
parameter $\lambda(Q^2)$ of $F_2(x, Q^2)\sim x^{-\lambda}$   
decreases linearly with $\ln Q^2$ down to $Q^2\approx 1$ GeV$^2$, 
flattening out and becoming consistent with $\epsilon=0.1$ only below 
$Q^2=1$ GeV$^2$. We therefore assume partonic behavior in diffractive
interactions and 
attempt to derive the parameters $\epsilon$ and
$\kappa$ from the nucleon pardon distribution functions (pdf's) 
at $Q^2=1$ GeV$^2$, shown in Fig.~\ref{fig:softQCD}~({\em right})
for the CTEQ5L parameterization.

The region of interest to diffraction, $x\leq 0.1$, is dominated by sea
gluons and quarks.
In this region, a fit of the form $xf(x)\sim x^{-\lambda}$ in 
Fig.~\ref{fig:softQCD}~({\em right}) yields 
$\lambda_g\approx 0.2$ and $\lambda_q\approx 0.04$ with relative weights 
$w_g\approx 0.75$ and $w_q\approx 0.25$~\footnote{For valence quarks
$\lambda\approx -0.5$; this is relevant for Reggeon contributions, which 
are not considered here.}. Noting that 
the number of wee partons grows as $\int_{1/s}^1 f(x)dx\sim s^\lambda$, 
the Pomeron intercept may be obtained from the parameters $\lambda_g$ and 
$\lambda_q$, appropriately weighted by a procedure involving 
gluon and quark color factors,
\begin{equation}
c_g=\frac{1}{N_c^2-1},\;\;\;\;c_q=\frac{1}{N_c}
\label{eq:color}
\end{equation}       
Weighting places $\epsilon$ in the range $\lambda_q<\epsilon<\lambda_g$, or $0.04<\epsilon<0.2$, 
which covers the experimental value of $\epsilon=0.104$. 
The  parameter $\kappa$ is obtained from the $g/q$ color factors and weights:
\begin{eqnarray}
\kappa\approx &c_gw_g+c_qw_q&=0.18
\label{eq:k_th}
\end{eqnarray}
This prediction is in remarkably good agreement with 
$\kappa_{exp}=0.17\pm0.02$.  
\section{Soft diffraction summary}
Using the (re)normalized gap probability model, which is based on a 
partonic description of diffraction, soft diffraction cross sections may 
be derived from the inclusive nucleon 
pdf's at $Q^2=1$ GeV$^2$. No free parameters 
are required in the formulation of the $t=0$ differential cross sections:
both the Pomeron intercept and the triple-Pomeron coupling 
(related to $\kappa$) are derived from 
the underlying inclusive pdf's. Except for normalization, the procedure 
outlined reproduces the Regge theory predictions. Renormalization correctly 
predicts the data for all single and double gap processes studied by CDF,
which include SD, DD, SDD and DPE. In all cases renormalization removes the 
explicit $s^{2\epsilon}$ dependence from the cross section expressions 
in terms of $M^2$. This leads to $M^2$-scaling, which appears to be 
a basic property of diffraction at high energies. In the parton model, 
$M^2$-scaling is traced back to the 
power law behavior of the inclusive pdf's at low $x$. In summary, 
diffraction appears to be a low-$x$ interaction subject to color constraints.
 
In Ref.\cite{R} it was pointed out that the Pomeron flux should be 
renormalized to unity only at high enough energies where it  
exceed unity. This remains true for central and multigap processes as well:
renormalization of the Regge gap probability should be performed only  
for $s$-values for which $P_{gap}(s)>1$. The parton model approach 
offers a simple explanation for this condition in terms of saturation effects. 

\section{Hard diffraction}
Hard diffraction processes are defined as those in which there is a hard 
partonic scattering in addition to the diffractive rapidity gap 
signature. Events may have forward, central, or multiple rapidity 
gaps, as shown in Fig.~(\ref{fig:hard})~({\em left}) 
for dijet production in $\bar pp$ collisions at the Tevatron, 
and  ({\em right}) for diffractive deep inelastic scattering (DIS) 
in $ep$ collisions at HERA. 

\begin{minipage}[t]{8.5cm}
\includegraphics{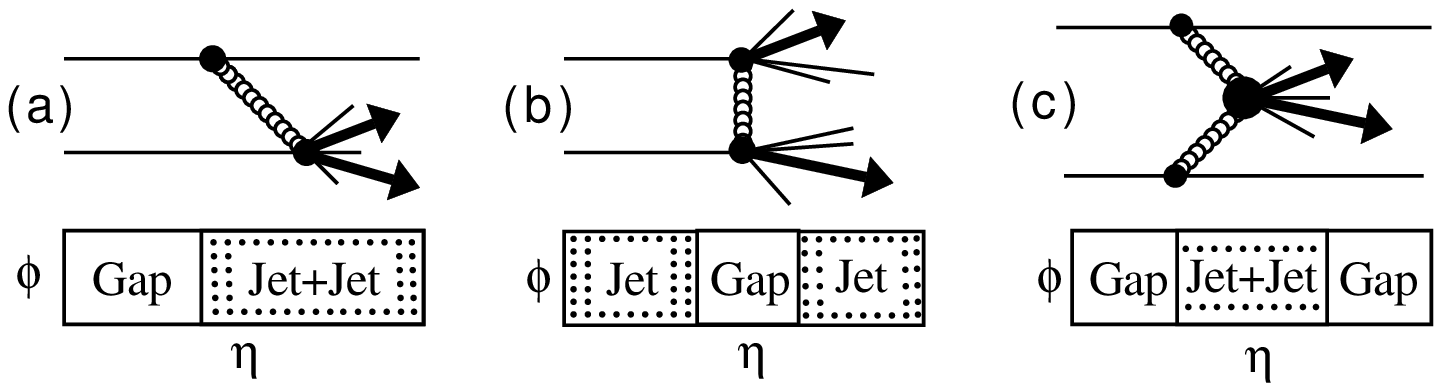}
\hfill
\end{minipage}
\begin{minipage}[t]{2.5cm}
\includegraphics{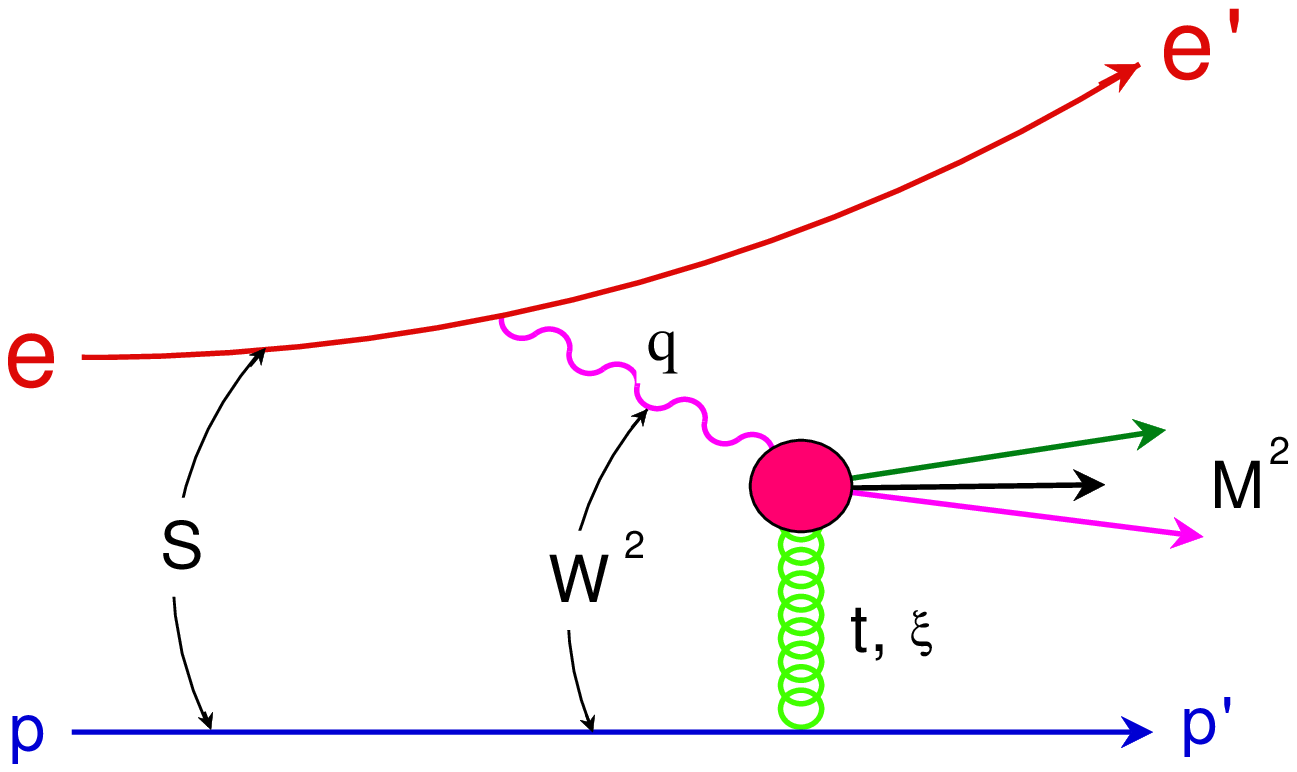}
\end{minipage}
\vglue 2cm
\begin{center}
\begin{figure}[htp]
\caption{\it  ({left}) Dijet production diagrams and event topologies for
$\bar pp$ (a) single diffraction (b) double diffraction 
and (c) double Pomeron exchange;
({right}) diffractive photon dissociation in $ep$ collisions.} 
\label{fig:hard}
\end{figure}
\end{center}

Results from Tevatron and HERA experiments have addressed several aspects 
of hard diffraction that are important for understanding its 
QCD nature. Process  dependence of diffractive fractions and detailed 
comparisons of diffractive structure functions between Tevatron and 
HERA and within Tevatron data point to a picture in which 
the generic exchange mediating diffraction, generally referred to as 
Pomeron, appears to be a hard scale gluon or quark color-shielded by 
other gluons/quarks from the soft sector of the proton structure.  
Saturation effects in the soft sector, which are prominent 
at the Tevatron, lead to breakdown of QCD factorization between 
HERA and Tevatron,  as well as within Tevatron data as 
a function of $\sqrt s$. 
Below, we briefly present some relevant results reported 
by the CDF Collaboration, compare Tevatron diffractive 
structure functions with predictions based on HERA data, and
examine the applicability of the renormalized gap probability parton model
approach to hard diffraction.   
\subsection{Process dependence}
Diffractive fractions have been measured by CDF for $W$\cite{CDF_W}, 
dijet\cite{CDF_JJG}, $b$-quark\cite{CDF_b}
and $J/\psi$\cite{CDF_jpsi} production at $\sqrt s=1800$ GeV, 
as well as for events with a gap between jets at 
$\sqrt s=1800$\cite{CDF_JGJ1800}  and $\sqrt s=630$\cite{CDF_JGJ630} GeV. 
Table~(\ref{tab:fraction}) presents the measured diffractive fractions 
within the indicated kinematic regions.

\begin{table}[ht]
\label{tab:fraction}
\begin{center}
\caption{\it  Diffractive fractions for forward and central gap processes at CDF}
\vglue 1em
\begin{tabular}{|l|c|c|c|}
\hline
Hard process&$\sqrt{s}$ (GeV)&$R=\frac{\rm DIFF}{\rm TOTAL}\,(\%)$&
Kinematic region\\
\hline\hline
$W(\rightarrow e\nu)$+G&1800&$1.15\pm 0.55$&$E_T^e,\;/\!\!\!\!E_T>20$ GeV\\
Jet+Jet+G&1800&$0.75\pm 0.1$&$E_T^{jet}>20$ GeV, $\eta^{jet}>1.8$\\
$b(\rightarrow e+X)$+G&1800&$0.62\pm 0.25$&$|\eta^e|<1.1$, $p_T^e>9.5$ GeV\\
$J/\psi(\rightarrow \mu\mu)$+G&1800&$1.45\pm 0.25$
&$|\eta^{\mu}|<0.6$, $p_T^{\mu}>2$ GeV\\
\hline
Jet-G-Jet&1800&$1.13\pm 0.16$&$E_T^{jet}>20$ GeV, $\eta^{jet}>1.8$\\
Jet-G-Jet&630&$2.7\pm 0.9$&$E_T^{jet}>8$ GeV, $\eta^{jet}>1.8$\\
\hline
\end{tabular}
\end{center}
\end{table}

The diffractive fractions at $\sqrt s=1800$ GeV are approximately 1\%. 
As the processes presented here have different sensitivities to the 
quark/gluon partonic component of the exchange, 
the measured fractions were used by CDF to extract the  
gluon fraction of the Pomeron~\cite{CDF_b}. 
The result is compared below with the gluon fraction obtained 
by H1 from diffractive DIS at HERA~\cite{heraglue}:
\begin{equation}
\hbox{ gluon fraction of $\pom$:}\;\;\;\;\;\;f_g^{\rm CDF}=0.54\pm 0.15\;\;\;\;
f_g^{\rm H1}=0.75\pm0.15 
\label{eq:glue}
\end{equation} 
The gluon fraction appears to be larger at HERA than at the Tevatron.
Another interesting result is that the Jet-G-Jet fraction is larger at 
$\sqrt s=630$ than at 1800 GeV by a factor of $2.3\pm0.9$.  
These results will be revisited below when we present a parton model 
approach for hard diffraction, in which diffractive fractions are 
controlled by the underlying inclusive pdf's subject to 
renormalization and color constraints.

\section{Factorization breakdown}
The success of QCD in describing hard scattering processes 
rests on the factorization theorem, which allows hadronic cross sections 
to be expressed in terms of parton-level cross sections convoluted 
with uniquely defined hadron partonic densities.
This success was shaken by the discovery that factorization breaks down
in diffractive processes when comparing $\bar pp$ collider results with 
results from diffractive DIS at HERA. The first comparison involved 
diffractive dijet production rates reported by UA8~\cite{UA8} 
and was made by this author in the paper proposing Pomeron flux
renormalization mentioned in section~\ref{sec:renorm}. 
In the same paper, predictions were made 
for the Tevatron, which were later confirmed by 
experiment~\cite{CDF_W,CDF_JJG,CDF_b,CDF_jpsi}.
However, these experiments involved cross sections integrated over 
a range of $x$-Bjorken ($x_{Bj}$ or $x$) and thus provided no 
$x$ or $Q^2$ dependence of the factorization breakdown. 
Such information came later from CDF measurements~\cite{CDF_jj1800,CDF_jj630} 
using a Roman Pot 
Spectrometer (RPS) to detect the leading antiproton in coincidence 
with two jets produced in the main detector,
\begin{equation}
\hbox{ SD dijet production:}\;\;\;\;\bar p+p\rightarrow \bar p+Jet_1+Jet_2+X
\label{diffdijet}
\end{equation}
Obtaining $x_{\bar p}$ from dijet production kinematics, CDF 
measured the ratio of rates of single diffractive (SD) 
to non-diffractive (ND) dijet production
\footnote{As this ratio is $<1\%$, we use `ND' and `inclusive' rates 
interchangeably throughout this paper.}, 
which in LO QCD approximately equals the ratio 
of the corresponding structure functions. 
From the ratio of rates, and after changing 
variables from $x$ to $\beta$, where $\beta\equiv x/\xi$ is the fractional 
momentum of the struck parton in the Pomeron, the diffractive structure 
function was obtained as a function of $\beta$ and compared with 
predictions based on diffractive pdf's from HERA. 

The structure function 
measured in dijet production has contributions from gluon and quark 
pdf's weighted by appropriate color factors:
\begin{equation}
F_{jj}(x)=x\left[g(x)+\frac{4}{9}q(x)\right]
\label{eq:SF}
\end{equation} 
The CDF results for $R(x)$, the SD/ND ratio as a function of $x$, and for 
$F_{jj}^D(\beta)$ versus $\beta$ are presented in 
Fig.~\ref{fig:SF} ({\em left}) and ({\em right}), respectively.

\begin{minipage}[t]{5.5cm}
\includegraphics{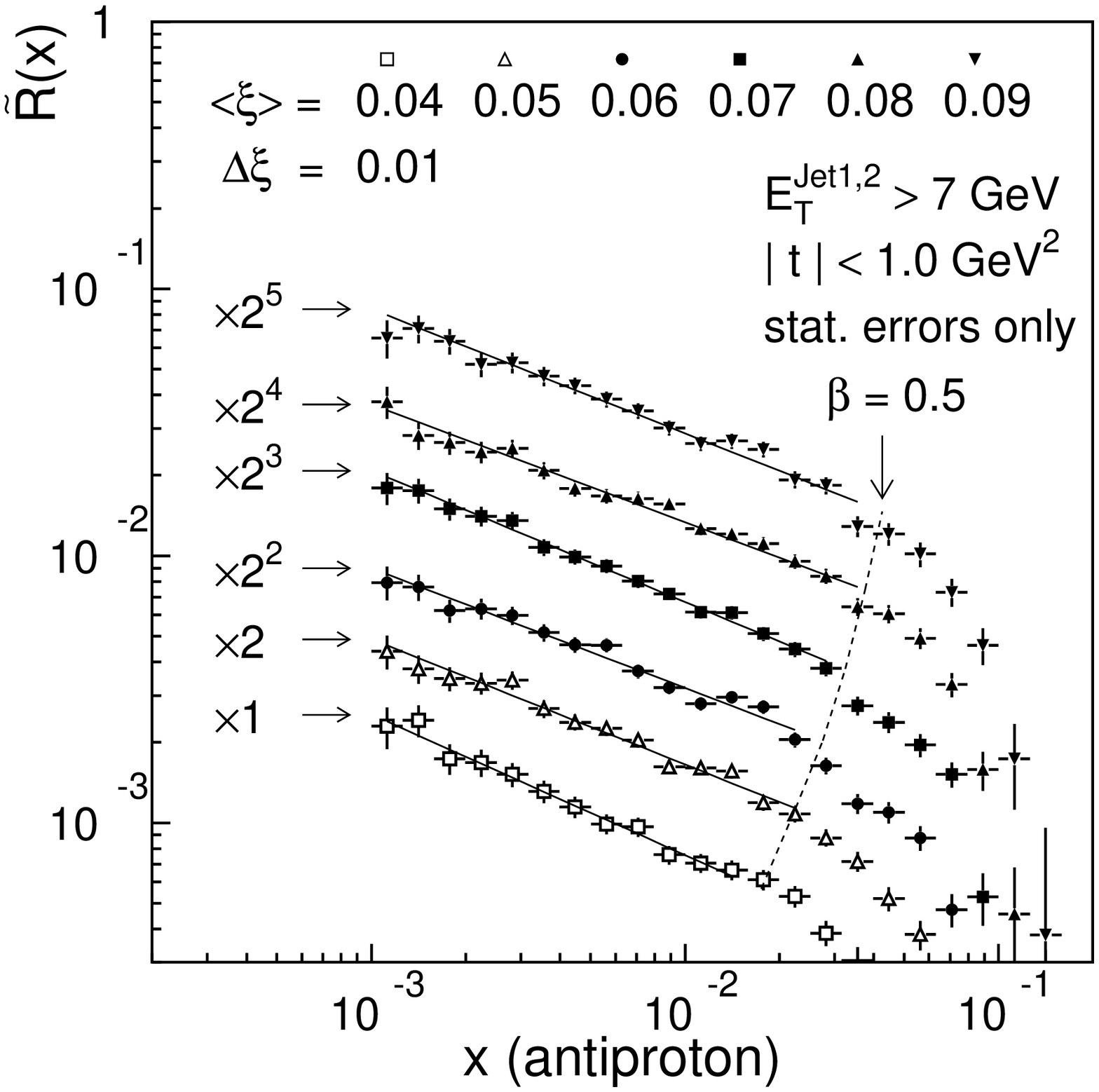}
\hfill
\end{minipage}
\begin{minipage}[t]{5.5cm}
\includegraphics{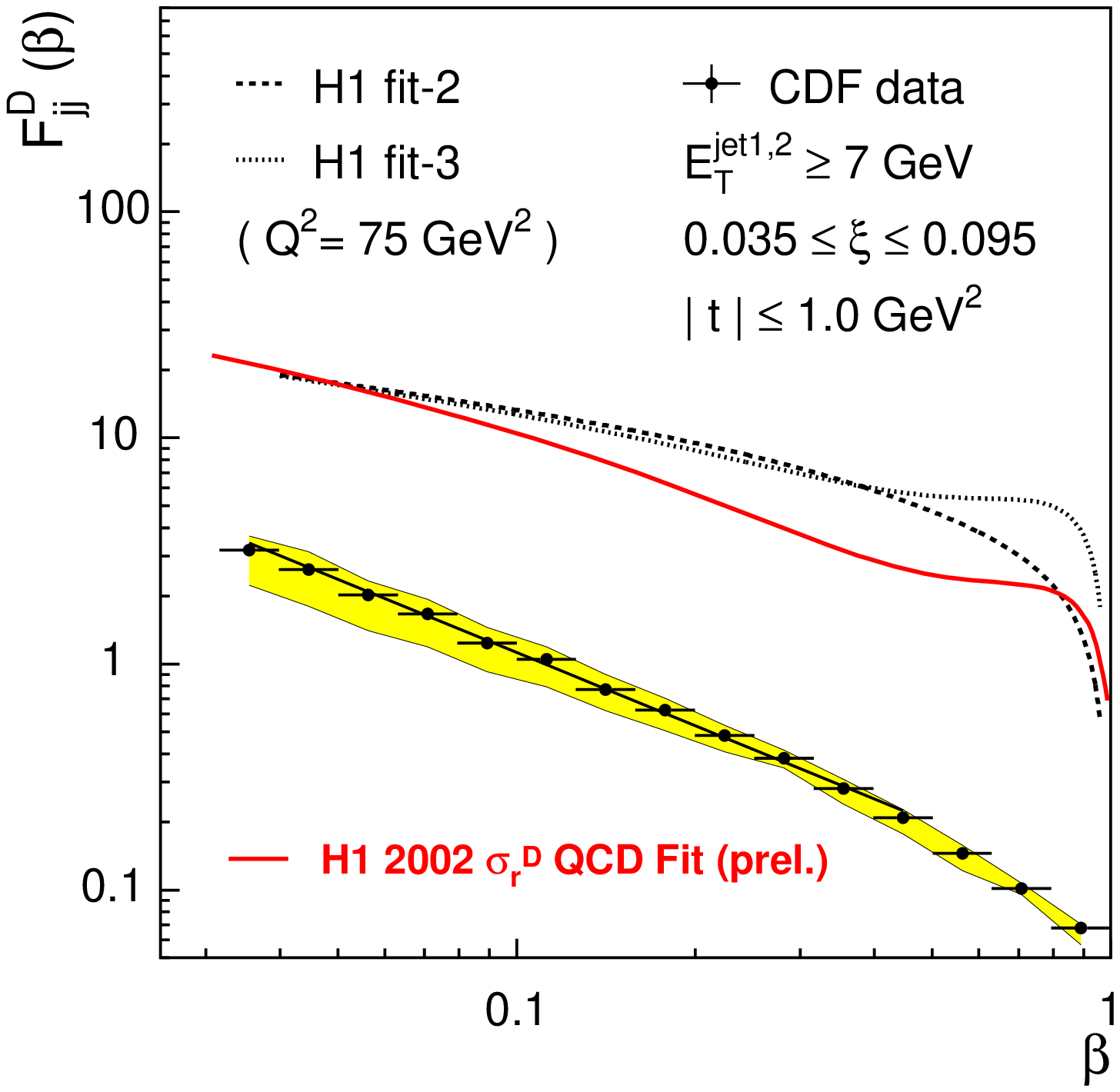}
\end{minipage}
\vglue 5cm
\begin{figure}[h]
\caption{\it  CDF diffractive dijet data at $\sqrt s=1800$ GeV:
(left)\, Ratio of SD to ND dijet production rates versus $x_{\bar p}$;
(right)\, the CDF diffractive structure function versus $\beta$ compared with 
predictions based on factorization and parton densities obtained 
by H1 from diffractive DIS at HERA~\cite{heraglue}.}
\label{fig:SF}
\end{figure}  
The data exhibit the following interesting features:
\begin{itemize}
\item Fig.~\ref{fig:SF}~{(\em left)}\,: In the region of $0.035<\xi<0.095$ and 
$\beta\equiv x/\xi<0.5$, the ratio of SD to ND rates (or structure functions) 
has no significant $\xi$ dependence and decreases 
with increasing $x$ as
\begin{equation}
R^{SD/ND}(x)=R_0\cdot x^{-r};\;\;R_0=(6.1\pm0.1)\times 10^{-3},\;\; r=0.45\pm0.02
\label{eq:SFratio}
\end{equation}
\item  Fig.~\ref{fig:SF}~{\em (right)}\,: For $\beta<0.5$, $F_{jj}^D(\beta)$ 
agrees in shape with the prediction from HERA but is suppressed 
by a factor of $\approx 10$ (comparisons for $\beta>0.5$ 
should be avoided due to large systematic errors in the prediction).
\end{itemize}
The breakdown of factorization between Tevatron and HERA is generally 
attributed to additional partonic interactions that spoil the rapidity 
gap formed by Pomeron exchange. Under this scenario, factorization should 
also break down in diffractive $\bar pp$ collisions at the Tevatron
between different collision energies or between single and double gap 
processes.

\subsection{$s$-dependence of $F^D_{jj}$}
Since the number of partons gets smaller at lower $s$, $F^D_{jj}$ 
should increase as $s$ decreases. CDF measured 
the ratio of diffractive dijet production at $\sqrt s=630$ over 
1800 GeV to be $1.3\pm 0.2{(\rm stat)}^{+0.4}_{-0.3}{\rm (syst)}$~\cite{CDF_jj630}, 
but due to the large uncertainties no definitive conclusions can be drawn 
about the $s$-dependence of the factorization breakdown.

\subsection{$F^D_{jj}$ from double gap events}
Factorization implies that the double-ratio $D$  
of dijet production in SD over ND events, $R^{SD}_{ND}$,   
to that of DPE over SD, $R^{DPE}_{SD}$, should be unity. 
However, since particles produced by additional partonic interactions 
spread throughout the entire available rapidity region, the two gaps 
in a DPE event either both survive or are simultaneously spoiled. 
Thus, factorization is expected to break down between the above 
two ratios, resulting in a deviation of $D$ from unity.
A CDF measurement of dijet production in DPE events
yielded $D=0.19\pm0.07$~\cite{CDF_DPE}, 
confirming the expectation that the formation 
of the second gap is not, if at all, suppressed.
This result is similar to the one discussed in section~\ref{sec:multigap}
for soft diffraction events.
 
\section{Restoring factorization}
For non-suppressed gaps, factorization should hold. This hypothesis was 
tested in a comparison~\footnote{The comparison was performed by the author and K. Hatakeyama 
using CDF published data~\cite{CDF_DPE} and preliminary 
H1 diffractive parton densities~\cite{heraglue}.}
of the diffractive structure function measured on the proton side 
in events with a leading antiproton at the Tevatron with 
expectations from diffractive DIS at HERA. Results are shown in 
Fig.~(\ref{fig:restor}). The approximate agreement between Tevatron data and 
the expectation from HERA shows that factorization is largely 
restored for events that already have a rapidity gap. 
\vglue -1em
\begin{minipage}[t]{5.5cm}
\includegraphics{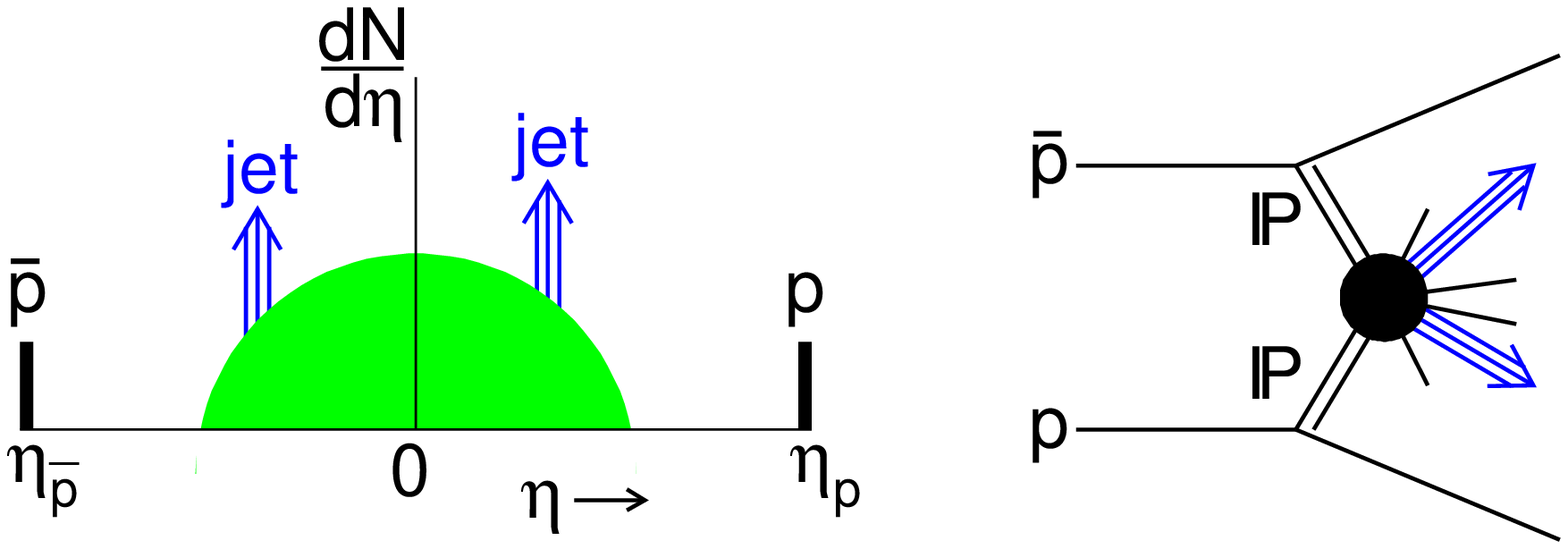}
\hfill
\end{minipage}
\begin{minipage}[t]{5.5cm}
\includegraphics{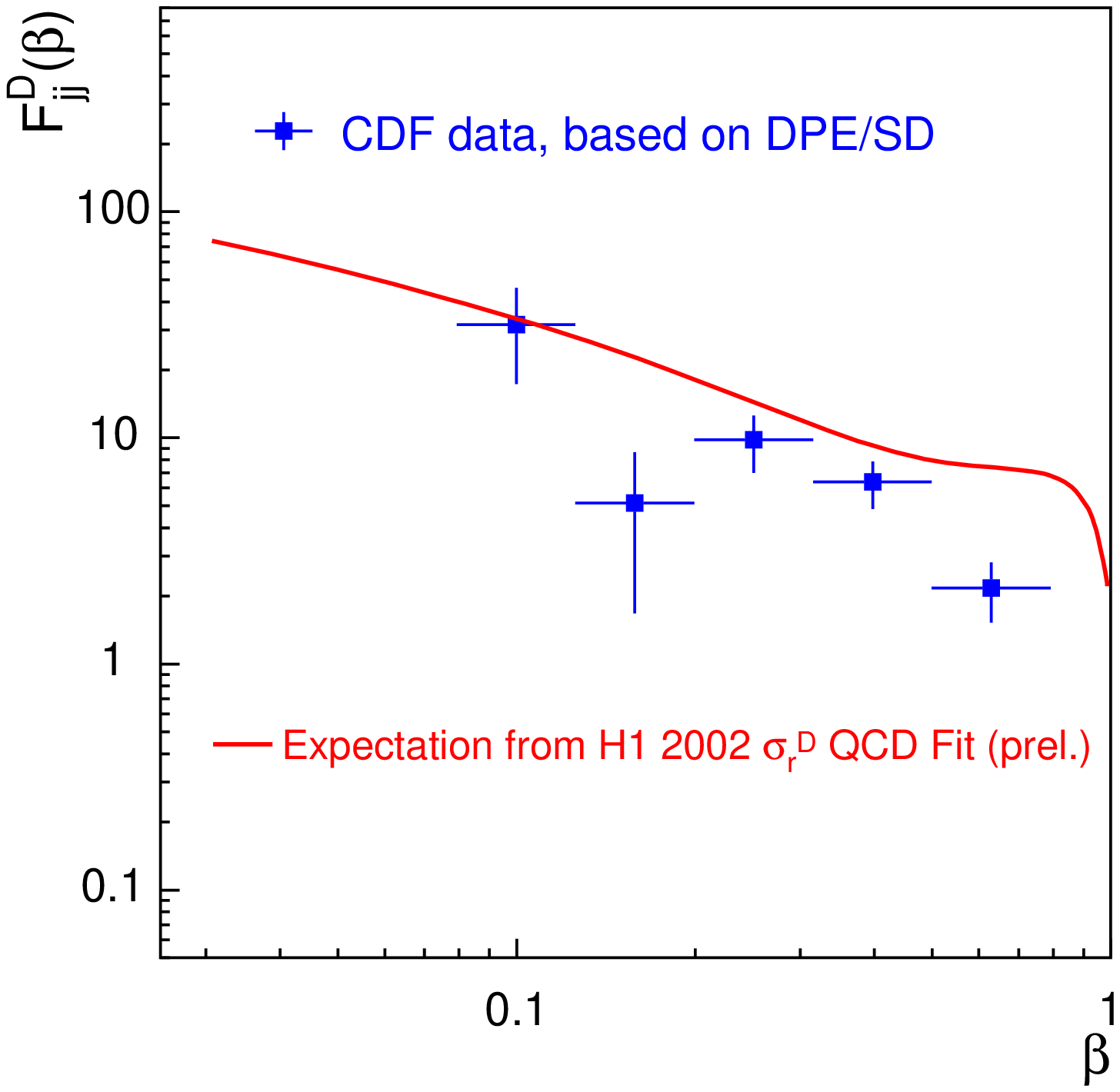}
\end{minipage}
\vglue 5cm
\begin{figure}[h]
\caption{\it (left)\, Schematic representation of dijet production in DPE; 
(right)\, comparison of $F_{jj}^D(\beta)$ of the proton in DPE events 
with expectations from H1 parton densities from diffractive DIS at HERA.}
\label{fig:restor}
\end{figure}

\section{Rapidity gap survival probability}
Hard diffraction calculations invariably involve
estimates of the gap survival probability~\cite{Bj} 
based on renormalization~\cite{R}, 
``screening''~\cite{GLM,KMR}, or color neutralization 
by soft color exchanges~\cite{softcolor}. The non-perturbative nature of 
these estimates makes them model-dependent and difficult to 
obtain for all processes across the soft and hard diffractive sectors.
The CDF two-gap results  
indicate that calculations for diffractive processes which already have a 
rapidity gap are largely free from gap survival considerations. Thus, 
multigap diffraction   
opens the way for QCD studies of diffraction without the complications arising
from gap survival and will undoubtedly become the arena on which 
diffraction will be confronted by QCD at the Large Hadron Collider.

\section{Hard diffraction in a QCD framework}
\begin{center}
\begin{figure}[htp]
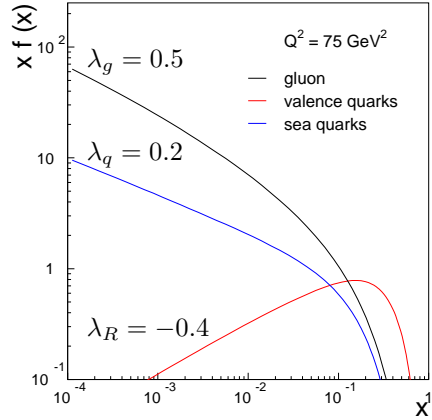

\caption{\it CTEQ5L nucleon parton distribution functions for $Q^2=75$ GeV$^2$. The parameters $\lambda_{g,q,R}$ are the slopes of the gluon, 
sea quark, and valence quark distribution (`$R$' stands for $Reggeon$) 
in the region of $x<0.1$ where the power law 
behavior holds.} 
\label{fig:hardQCD}
\end{figure}
\end{center}

In this section we adapt the QCD approach of section~\ref{approach}
to diffractive DIS at HERA and diffractive dijet 
production at the Tevatron:
\begin{eqnarray}
\hbox{ HERA:}&&\gamma*+p\rightarrow p+Jet+X\nonumber\\
\hbox{ Tevatron:}&&\bar p+p\rightarrow+\hbox{dijet}+X\nonumber
\label{SD_hera_Tev}
\end{eqnarray}
The hard process may involve several color ``emissions''
from the surviving proton comprising a color singlet with vacuum 
quantum numbers. Two of the emissions have special importance: 
the one at $x=x_{Bj}$ from the proton's hard pdf at scale $Q^2$, which is responsible 
for the hard scattering, and the other at $x=\xi$ from the soft 
pdf at $Q^2\approx 1$ GeV$^2$, which neutralizes the exchanged color 
and forms the rapidity gap. The diffractive structure function should 
then be~\cite{brown} the product of the inclusive 
structure function and the soft parton 
density at $x=\xi$ [for simplicity, 
we do not include $t$ dependence],
\begin{equation}
F^D(\xi,x,Q^2)\propto \frac{1}{\xi^{1+\epsilon}}\cdot F(x,Q^2)
\sim\frac{1}{\xi^{1+\epsilon}}\cdot
\frac{C(Q^2)}{(\beta\xi)^{\lambda(Q^2)}}
\Rightarrow
\frac{A_{\rm norm}}{\xi^{1+\epsilon+\lambda}}
\cdot c_{g,q}\frac{C}{\beta^{\lambda }}
\label{eq:F2D3}
\end{equation}
where $c_{q,q}$ are the color factors shown in Eq.~(\ref{eq:color}), 
$\lambda$ is the parameter of a power law fit to the relevant hard structure 
function in the region $x<1$ (see Fig.~\ref{fig:hardQCD}),
and $A_{\rm norm}$ is a normalization factor.

\begin{minipage}[t]{5.5cm}
\vglue 2.5cm
\hspace{2cm}{\large $x\cdot f(x)\propto x^{-\lambda}$}
\end{minipage}
\begin{minipage}[t]{6.5cm}
\includegraphics{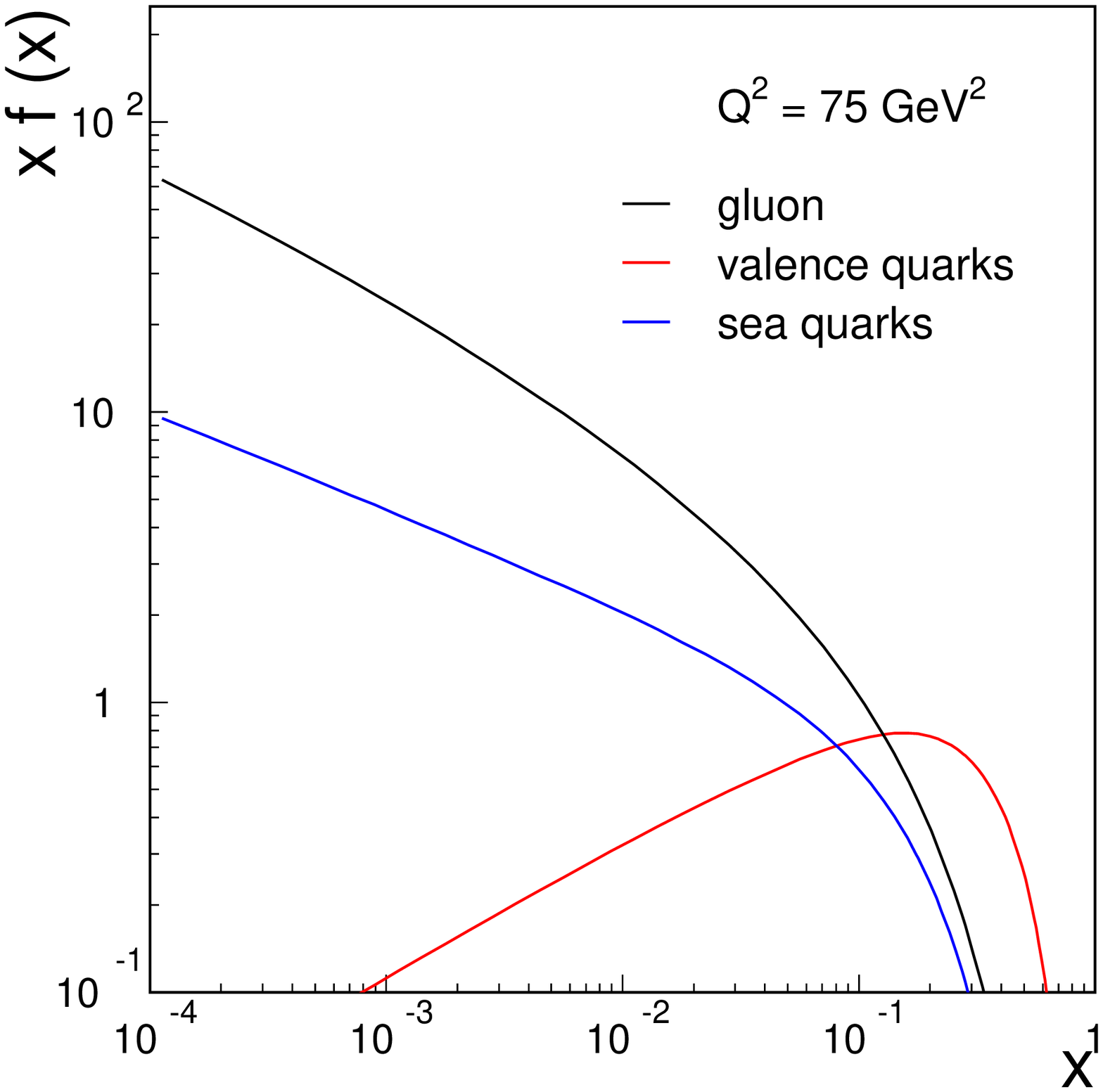}
\vglue 0.7cm
\hspace{1.8cm}\hspace{1cm}$\lambda_g=0.5$\\
\vglue 0.4cm
\hspace {1.8cm}$\hspace{1cm}\lambda_q=0.2$\\
\vglue 1.5cm
\hspace {1.8cm}$\hspace{1cm}\lambda_R=-0.4$\\
\end{minipage}

\subsection{Predictions for the diffractive structure function at HERA}
At high $Q^2$ at HERA, where factorization is expected to 
hold~\cite{R,JCollins}, $A_{\rm norm}$ is simply the (constant) normalization 
factor of the soft pdf. This constant normalization leads to two 
important predictions:
\begin{itemize}
\item The Pomeron intercept in diffractive DIS (DDIS) 
is $Q^2$-dependent and equals 
the average of the soft and hard intercepts:
\begin{eqnarray}
\alpha^{DIS}_{\pom}&=&1+\lambda(Q^2)\\
\alpha^{DDIS}_{\pom}&=&1+\frac{1}{2}\left[\epsilon+\lambda(Q^2)\right]\nonumber
\label{eq:heraintercept}
\end{eqnarray}
\item The ratio of DDIS to DIS structure functions at a given $\xi$ 
is independent of $x$ and $Q^2$:
\begin{equation}
R\left[\frac{F^D(\xi,x,Q^2)}{F^{ND}(x,Q^2)}\right]_{\rm HERA}=
\frac{A_{\rm norm}\cdot c_q}{\xi^{1+\epsilon}}=
\frac{\rm constant}{\xi^{1+\epsilon}}
\label{eq:Rhera}
\end{equation}
\end{itemize}
HERA data are consistent with these predictions~\cite{heraglue}.

\subsection{Predictions for the diffractive structure function at the Tevatron}
At the Tevatron, where high soft parton densities lead to saturation,
$A_{\rm norm}$ must be renormalized $ala$ Eq.~(\ref{eq:renorm}) by 
being replaced by
\begin{equation}
\hbox{ Tevatron:}\;\;\;A_{\rm renorm}=1/\int^{\xi=0.1}_{\xi_{min}}
\propto
\left(\frac{1}{\beta s}\right)^{\epsilon+\lambda}
\label{eq:tevrenorm}
\end{equation}
where we have used $\xi_{min}=x_{min}/\beta$ and $x_{min}\propto 1/s$.
Thus, the diffractive structure function acquires a term
$\sim (1/\beta)^{\epsilon+\lambda}$, and the diffractive to inclusive 
structure function ratio a term $\sim (1/x)^{\epsilon+\lambda}$.
This prediction is confirmed by the CDF data on dijet production,
where the $x$-dependence of the diffractive to inclusive ratio 
was measured to be $\sim 1/x^{0.45}$ (see Eq.~\ref{eq:SFratio}) 
\footnote{In calculating the value $r$ of the 
$\sim 1/x^r$ dependence, care should be 
taken to separately renormalize the sea gluon and sea quark contributions,  
and also consider the contribution of the valence quarks (Reggeon 
contributions).}. 

\section{Summary and conclusion}
We have reviewed experimental data on soft and hard diffraction, 
concetrating on aspects that point to the partonic nature of the  
exchange (Pomeron) responsible for rapidity gap formation. 

In soft diffraction, the exponential dependence of total [elastic] 
cross sections on $\Delta y'$ [$\Delta y$], the rapidity region 
in which there is [there is not] particle production, allows  
differential diffractive cross sections to be factorized into two terms:
one representing the total cross section at the reduced energy squared,
defined by $\ln (s'/s_0)=\sum_i\Delta y_i'$, and the other depending 
on $\sum_i\Delta y_i$ and interpreted as the gap formation probablity.
By (re)normalizing the latter to unity and multiplying by a color factor
$\kappa$ ($\kappa^n$ for $n$ gaps) 
derived from gluon and quark color factors weighted by the corresponding  
inclusive pdf's at $Q^2\approx 1$ GeV$^2$, 
cross sections for single, central, and multigap diffraction are obtained,
which are in excellent agreement with experimental results.
Prominent aspects of the data explained in this model are the
$M^2$-scaling behavior and the independece on the number of
gaps of the suppression of diffractice cross sections at high 
energies relative to Regge theory predictions.

In hard diffraction, the diffractive structure function is obtained 
by convoluting the inclusive $F(Q^2,x)$ with the parton density at 
$x=\xi$ at scale $Q^2\approx 1$ GeV$^2$, identifying the 
gap formation probability term, and (re)normalizing the latter to unity at
the Tevatron. Aspects of the data explained by this approach are 
the rise of the Pomeron intercept with $Q^2$ and the constant DDIS/DIS 
ratio versus $x$ and $Q^2$ at HERA, as well as the falling 
as $\sim x^{-r}$ of the SD/ND ratio at the Tevatron. Also explained 
in this model is the factorization breakdown (and restoration for 
multigap diffraction!)  between Tevatron diffracive 
structure functions and expectations from DDIS at HERA. 
 
In conclusion, scaling and factorization aspects of the data have been  
phenomenologically interpreted in a parton model approach, in which 
diffraction is treated as a low $x$ partonic interaction subject 
to color constraints.

\end{document}